\documentclass{aa}
\usepackage{psfig,times}
\newcommand{\mkn}{\hbox{Mrk~421} }
\newcommand{\xmm}{\hbox{XMM$-$Newton} }
\newcommand{\etal}{et al. }
\def\ltsima{$\; \buildrel < \over \sim \;$}
\def\simlt{\lower.5ex\hbox{\ltsima}}            
\def\gtsima{$\; \buildrel > \over \sim \;$}
\def\simgt{\lower.5ex\hbox{\gtsima}}            

\def\flux{erg\,cm$^{-2}\,s^{-1}$}

\begin{document}
\title{XMM$-$Newton observations of Markarian~421\thanks{Based on
    observations with XMM$-$Newton, an ESA Science Mission 
    with instruments and contributions directly funded by ESA Member
    States and the USA (NASA) }}
\author{W. Brinkmann\inst{1} \and S. Sembay\inst{2} \and R. G. Griffiths\inst{2}
 \and G. Branduardi-Raymont\inst{3} \and M. Gliozzi\inst{1} 
 \and Th. Boller\inst{1} \and  \\ A. Tiengo\inst{4}
 \and S. Molendi\inst{5} \and S. Zane\inst{3}  } 
\offprints{W. Brinkmann,  \email{wpb@mpe.mpg.de}}
\institute{ Max--Planck--Institut f\"ur extraterrestrische Physik,
   Giessenbachstrasse, D-85740 Garching, Germany
\and  X-ray Astronomy Group, Department of Physics and Astronomy,
  University of Leicester, LE1 7RH, U.K.
\and  Mullard Space Science Laboratory, University College London,
             Holmbury St Mary, Dorking, Surrey, RH5 6NT, U.K.
\and XMM-Newton Science Operation Centre, Villafranca Satellite Tracking
Station, 28080 Madrid, ES
\and Istituto di Fisica Cosmica "G.Occhialini", Via Bassini 15, I-20133, 
 Italy}
\date{Received  ? / Accepted ?}
 
\abstract{
The BL Lac object \mkn  was  observed on May 25, 2000 during the 
\xmm Cal/PV phase.
The high throughput of the X-ray telescopes and the spectral capabilities
of the instruments allow an uninterrupted temporal and spectral study of the 
source with unprecedented time resolution. \hfill \break
Mrk~421 was found at a relatively high state with a 2$-$6~keV flux of
$(1.3 - 1.9)\times10^{-10}$ erg\,cm$^{-2}\,s^{-1}$.
The observed intensity variations by more than a factor of three at highest
X-ray energies are accompanied by complex spectral variations 
with only a small time lag ($\tau = 265^{+116}_{-102}$ seconds)
between the hard and soft photons.\hfill \break
The (0.2$-$10)~keV spectrum can be well fitted by a broken power law  and 
no absorption structures
are found in the source spectrum at the high spectral resolution
of the transmission gratings.
}

\titlerunning{XMM observations of \mkn}
\authorrunning{W. Brinkmann et al.}
\maketitle
     
\section{INTRODUCTION}
\smallskip
Mrk~421 is the brightest BL Lac object at X-ray and UV wavelengths and it
is the first extragalactic source discovered at TeV energies (Punch \etal 
1992).
This nearby ($z = 0.031$) X-ray bright BL Lac has been observed by
 essentially all
previous X-ray missions and shows remarkable X-ray variability
correlated with strong activity at TeV energies (e.g., Takahashi \etal 1996,
Maraschi \etal 1999).
   
BL Lacs are thought to be dominated by relativistic jets seen at small
 angles to the line of sight (\cite{UP95}), 
and their radio-through-X-ray spectra are well fitted
by inhomogeneous jet models (\cite{BMU}).
However, the structure of the relativistic jets remains largely
unknown as the models are generally under-constrained by single epoch
 spectra and  the typical  smooth and nearly featureless blazar
spectra can be reproduced by models with widely different assumptions
(e.g., \cite{KOE}).    
  
Combining spectral and temporal information greatly constrains the 
jet physics.
Time scales are related to the crossing times of the emission regions
which depend on wavelength and/or the time scales of micro-physical processes
like acceleration and radiative losses.
The measured lags between the light curves at different energies as well as
spectral changes during intensity variations allow to probe the  micro-physics
of particle acceleration and radiation in the jet.
Thus \xmm with its  high sensitivity and broad 
energy bandwidth is an ideal tool to study BL Lacs 
as it allows spectroscopy with unprecedented time resolution,
uninterrupted by gaps because of the long period of the satellite orbit.
  
Mrk~421 was the first BL Lac object to be established as an X-ray source
(Ricketts \etal 1976, Cooke \etal 1978) and subsequent observations 
indicated that the X-ray spectrum has a soft power law form (Mushotzky
\etal 1978, Hall \etal 1981) which exhibits significant 
variability (Mushotzky \etal 1979).
More detailed studies with IUE and EXOSAT  showed that the variability
occurs on time scales of typically a day with an e-folding time scale
of $\sim 5\times10^4$ s (George \etal 1988). The source shows 
a dichotomy of X-ray states: a  low, soft state ($f_{2-6 {\rm keV}} \simlt
2\times10^{-11}$ \flux, $\Gamma \sim 2.8$) where 
the source hardens when it brightens and a hard
´outburst´ state  ($f_{2-6 {\rm keV}} \simgt 8\times10^{-11}$ \flux) 
during which the spectral index remains at $\Gamma \sim 2$. 
 
In several Ginga observations, partly simultaneously with ROSAT,
 \mkn  was found at intermediate fluxes  of 
 $f_{2-6~{\rm keV}} = (3.6 - 5.2)\times10^{-11}$ \flux 
(\cite{MAK92}, Tashiro 1994). 
The data indicated that the amplitude of the flux variations with time scales
of a few hours got larger with increasing energy and the 
correlation between flux and spectral index was inconsistent with
that observed by EXOSAT. The quality of the spectral fits improved
considerably by using a broken  power law or a power law with
exponential cut off and the Ginga spectrum was significantly
steeper than the simultaneous ROSAT spectrum.
   
Since its discovery as a TeV source several multi-wavelength campaigns
have been conducted to study possible time lags between the X-ray band
and TeV energies and to investigate the pronounced spectral evolution
during flares seen in  X-rays  with ASCA  and BeppoSAX
 (Macomb \etal 1995, 1996,
Takahashi \etal 1996, Fossati \etal 1998, \cite{MA99}).
The source generally shows a complex behavior. While 
Takahashi \etal (1996) found a lag of about 4000 seconds between the soft
(0.5 - 1.0~keV) photons and the hard band (2 - 7.5~keV),
which was interpreted as an effect of radiative cooling, recent ASCA
observations show both, positive and negative lags (Takahashi \etal 2000).
BeppoSAX observations of a flare in April 1998, simultaneously observed
at TeV energies, showed that the hard photons lag the soft ones by
2-3 ksec and that, while the light curve is symmetric at softest X-ray energies,
it becomes increasingly asymmetric at higher energies with the 
decay being slower than the rise (Fossati \etal 2000).
  
Fitting the ASCA data by a simple power law Takahashi \etal (1996) find
that an absorbing column density considerably higher than the Galactic
value of $N_H = 1.5 \times 10^{20}$ cm$^{-2}$ (Elvis \etal 1989) 
is required to obtain
acceptable fits. Fixing the absorption at the Galactic value  a 
broken power law model provides a better fit than a simple power
law, but the $\chi^2_{\rm red}$ is often un-acceptable. With these
models the break energy is at $\sim$ 1.5 keV, and the change of the 
power law index at the break point is $\Delta \Gamma \sim 0.5$.
 
With the wider energy range of BeppoSAX it became clear that these simple
models are not adequate descriptions of the downward curved Synchrotron spectra
(Fossati \etal 2000) and continuously curved shapes had to be employed
(Inoue \& Takahara 1996, Tavecchio \etal 1998).
The Synchrotron peak energy varied between 0.4$-$ 1~keV, the spectral
index at an energy of 5~keV between $1.5 \leq \alpha \leq 2.2$.
Both quantities are correlated with the X-ray flux: the peak energy
positively, the spectral slope inversely:  with increasing
flux the synchrotron peak shifts to higher energies and the spectrum at
5~keV gets flatter.

Most of these results were obtained from data integrated over 
wide time intervals (typically one satellite orbit)  and from
giant flares with time scales of a day.
Uninterrupted data with high temporal and spectral resolution
can only be provided by  \xmm with its high sensitivity,
spectral resolving power, and broad energy band.
  
\begin{figure}
\psfig{figure=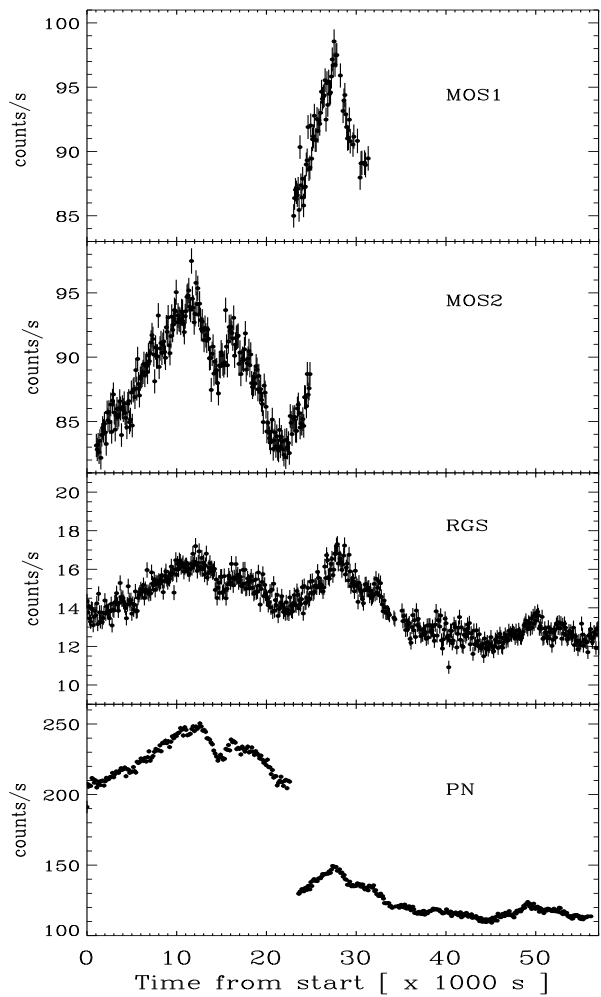,height=13.5truecm,width=8.5truecm,angle=0,%
  bbllx=120pt,bblly=70pt,bburx=440pt,bbury=730pt,clip=}
\caption[]{\small Light curves of \mkn during the \xmm observation
  in the different instruments. 
\label{figure:lctot}}
\end{figure}

\section{\xmm observations}
\smallskip
Mrk~421  was observed with the European Photon Imaging Camera (EPIC) and
the RGS  during orbit  84 
(May 25, 2000) of the calibration and performance verification phase
(Cal/PV) of XMM$-$Newton. The EPIC-PN camera was first operated from
UT 3:53 - UT 10:11 in the fast timing mode, then, from UT 11:34 to UT 19:34, in
Small Window  (SW) mode.
As the data analysis tools for the timing mode are not yet fully established
we will restrict our discussion mainly to the second part of the observation,
where the PN camera was operated in the SW mode.
This mode, with a frame integration time of $\sim$ 5.7 msec was chosen to
avoid photon pile-up in this strong source. Overall, a total of more
than 8.9 million counts were accumulated in SW mode in a net observation time of
\simlt 33 ksec, predominantly originating from the source itself.
  
Two separate observations of Mrk~421 were performed with the MOS cameras,
one in timing mode and the other in partial window mode with a free
running readout (PRI PART W4). During both observations the medium filter
was selected. For the purposes of this paper, however, only data taken in 
partial window mode is used. In PRI PART W4 mode only a fraction of the
central chip (100 x 100 pixels) is read out giving a time resolution of
~0.2 s. Exposure times of 7 ksec and 24 ksec were obtained by MOS1 and MOS2
respectively. A standard reduction of raw MOS event lists was performed,
using the XMM$-$Newton  Science Analysis System (SAS) and only 'real'
X-ray events with a pattern 0-12 were used to create spectra and light
curves for further analysis.

MOS data and PN
imaging data only have a small overlap in time. Due to a radiation alert
the MOS cameras were switched off shortly after the PN SW mode 
 observation began and were not turned on again.
 So the $\sim$30 ksec of available MOS time largely
precedes the PN observation; the actual overlap is only about
5 ksec.
  
Both RGS instrument chains also observed Mrk~421 in two
exposures for a total observing time of 64 ksec; thus the RGS coverage is 
slightly longer than that of EPIC. The data were acquired in the RGS
standard spectroscopy mode, where the full spectrum is read-out every $\sim$
6 s (for details on the RGS instrument see den Herder et al., this volume).
The data were processed with the SAS,
using the most up-to-date calibration parameters.

Fig. 1 shows the light curves of the whole observation for the 
different instruments. 
The count rate plots provide only a qualitative picture
as the  photons in the individual instruments are selected according 
to different criteria, neither covering the same energy
ranges nor originating from identical spatial regions around the source. 
This is most clearly seen in the PN data taken in Fast Timing and  in
Small Window mode, which  do not match in count rate 
when the modes were changed.
The quantitatively correct conversion from count rates to fluxes
is detector and operation mode  dependent and must be deferred 
to a later paper. 
 
No scientific data are available from the
Optical Monitor for this observation of Mrk~421.
 
\begin{figure}
\psfig{figure=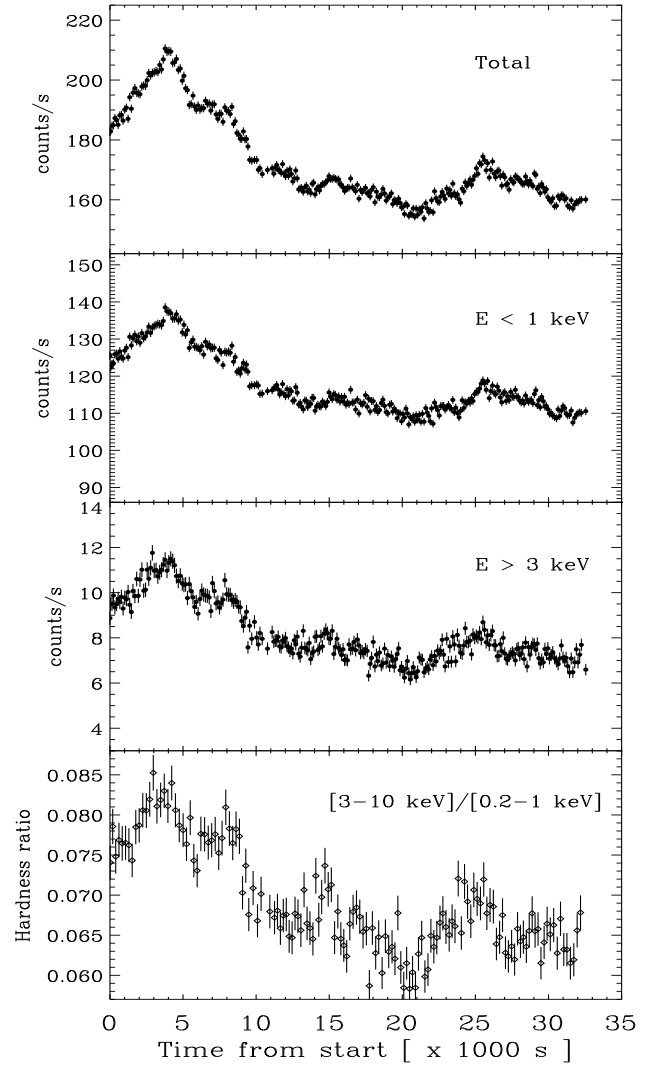,height=14.5truecm,width=8.5truecm,angle=0,%
bbllx=120pt,bblly=70pt,bburx=440pt,bbury=730pt,clip=}
\caption[]{\small The EPIC PN light curve of \mkn during orbit 84.
 The time is counted from the start of the SW mode observation. Only single
 events were taken and a  time binning of 100 seconds has been used.
 Shown from the top are: the total 0.2-10~keV count rate ; the count rate for
 the soft photons  E$\leq$ 1 keV,  for the
  hard photons E $>$ 3 keV, and at the bottom the hardness ratio
  between the hard and the soft photons.
\label{figure:lc_3}}
\end{figure}

\subsection{PN light curves}
Light curves were produced by using the PN photon event files produced 
by the SAS. Only single events with  energies
 0.2~keV  \simlt E \simlt 10~keV were selected  
from a circle of 9 detector pixels radius (corresponding to $\sim$ 38\arcsec) 
 around the source center;  
 the background was taken from the outer source-free
region of the detector and amounted to about  3\% 
of the source counts.

The photons were first binned into 10~sec intervals which turned 
 out to be necessary as the accepted time intervals are frequently
disrupted by short gaps  where the camera fell into counting mode.
 Finally, the binned data were summed up  into 
typically 100~s bins providing an excellent signal to noise ratio. 

Fig. ~\ref{figure:lc_3} shows the background 
subtracted light curves in three energy intervals: the total 0.2-10~keV
count rate at the top, then the soft (E $\leq$ 1~keV) and the
hard (E $>$ 3~keV) light curves in the middle, all in 100 s time bins 
and, at the bottom, the hardness ratios of the count rates
in the $(3 - 10~ {\rm keV}) / (0.2 - 1~{\rm keV})$ bands.
  
The total light curve shows intensity variations of about  a factor of 1.5; 
at high energies the maximal variation is much stronger, about  a factor
of  3.4 during the SW mode exposure.
The source gets harder when it flares, in accordance to previous observations.
The hard flux is considerably more variable than the soft flux as
quantified by  
the excess variance (Nandra \etal 1997):   for the soft band it is 
$(4.29\pm0.01)\cdot 10^{-3}$  while for the hard band it is
$(2.45\pm0.05)\cdot 10^{-2}$. 
   
The source was found in a relatively high state. From the spectral fit to
the total data set (see Sect. 3) we obtain an average 2$-$6~keV flux of
$(1.3 - 1.9)\times10^{-10}$ erg\,cm$^{-2}\,s^{-1}$ which is considerably
higher than the 'classical' soft state, but still lower than in the
April 1998 observation of BeppoSAX (Fossati \etal 2000).
  
\begin{figure}
\psfig{figure=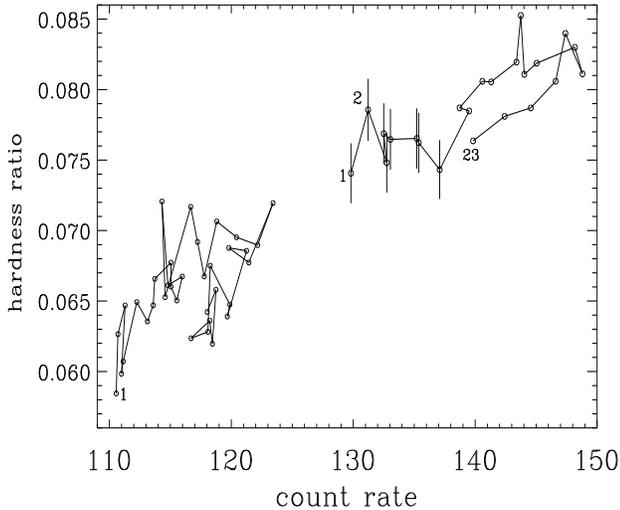,height=6.8truecm,width=8.5truecm,angle=0,%
  bbllx=50pt,bblly=365pt,bburx=550pt,bbury=700pt,clip=}
\caption[]{Spectral evolution during the two prominent flares  
 occuring at $\sim$ 3~ksec and $\sim$ 25~ksec after the start of the 
 observation. 
 Plotted are the hardness ratios as function of total count rate and
 a few error bars at the beginning of the first peak.
 Numbers indicate the temporal sequences of the data.} 
\label{figure:hardness}
\end{figure}

The light curve is rather complex and seems to consist of several,
partly overlapping 
flares on top of smooth intensity variations. The
spectral evolution during  the two prominent flares  at $\sim$ 3~ksec and
$\sim$ 25~ksec after the start of the observation shows a clockwise hysteresis,
as seen  in Fig. ~\ref{figure:hardness}.  This effect was 
already noted in previous ASCA observations (Takahashi \etal 1996) but 
as a property of the long term ($\sim$ 1 day) intensity variations of
the source.
   
The flares seem to have nearly linear rise- and decay profiles which
are asymmetric with energy dependent time scales. In the total band
the first flare rises slower ($\sim 4.7\times10^{-3}$ counts~s$^{-2}$)
than it decays ($\sim 8.8\times10^{-3}$ counts~s$^{-2}$), but in
general an exact determination of these time scales is problematic
due to the occurrence of additional, overlapping smaller flares.
In Fig. ~\ref{figure:hardness} these  lead  to a break of the 
clean intensity - hardness ratio correlation at the end of
the decaying part of the flares.
 
The flare profiles are energy dependent and we 
quantified the time delays between the hard and soft photons
employing the Z-transformed discrete cross-correlation function algorithm
(ZDCF) of Alexander (1997). The soft photons lag the
hard by $265^{+116}_{-102}$ seconds where the $1\sigma$ errors were
determined by using a Monte Carlo method described by Peterson \etal (1998).
Lags of a few 1000 secs as claimed from previous observations 
(Takahashi \etal 1996, Fossati \etal 2000) must thus be related to 
long-term flux changes and not to individual flares.
 
\subsection{Structure function analysis}
A structure function analysis is a method of
 quantifying time variability
without the problems encountered in the traditional Fourier
analysis technique in case of unevenly sampled data.
The first-order structure function (Simonetti \etal 1985) measures
the mean deviation for data points separated by a time
lag $\tau$, $SF(\tau)=\langle [F(t)-F(t+\tau)]^2\rangle$.
It is commonly
characterized in terms of its slope: $b=d \log(SF)/d\log\tau$.
One of the most useful features
of the structure function is its ability to discern the
range of time scales that contribute to the variations in the data set.
 For
lags shorter than the smallest correlation time scale and for lags longer
than the longest correlation time scale, the structure function displays two
plateau states ($b=0$) at different levels.
These regions are linked by a curve
whose slope depends on the nature of the intrinsic variation of the source
(e.g. flicker noise, shot noise, etc.).

\begin{figure}
\psfig{figure=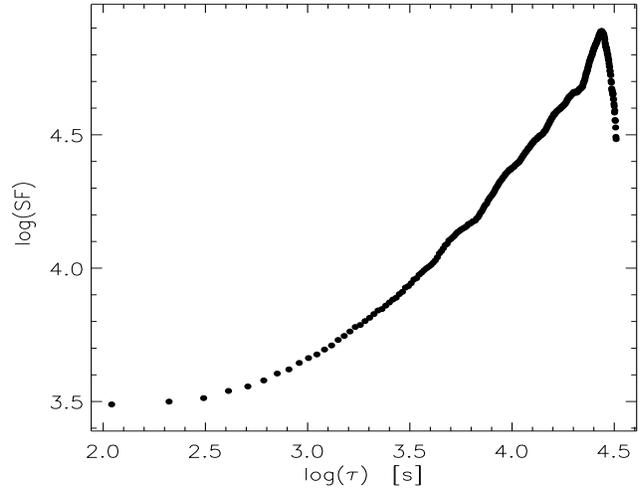,height=6.5truecm,width=8.5truecm,angle=0,%
  bbllx=59pt,bblly=180pt,bburx=515pt,bbury=598pt,clip=} 
\caption[]{\small Structure function of Mrk~421. Time lags are in secs.}
\label{figure:period}
\end{figure}

Fig. \ref{figure:period} shows the structure function of \mkn obtained from 
single events during the SW mode observation. The 'wavy' structure of the
increasing part must be related to the few individual flares in the light curve.
Mrk~421 shows little variation at time scales lower than
$10^3{~\rm s}$.
For longer time scales the slope of the structure function is
$b  \sim 1 $ which indicates  that  the nature of the variation
of the source can be ascribed to shot noise.
The current data do not allow to see the
roll-over at a lag of about half a day, found by ASCA (Takahashi \etal 2000).
The result is fully consistent with the findings of
Hughes \etal (1992), who obtained an average slope of $0.94\pm0.37$ 
for a sample of 20 BL Lac objects.
 
\section{Spectral analysis}
\medskip
The X-ray spectrum of \mkn was repeatedly studied in the past and 
fitted to a featureless power law or broken power law.
By taking the single events from the whole PN SW mode data set,
employing the latest versions of the detector response and CTE corrections, 
 we find that a power law with frozen Galactic absorption
 does not fit the 0.2-10~keV data  ($\chi^2_{red}$ = 2.025),
while  the fit improves leaving the absorption column density
free ($N_H = (3.05\pm0.07)\times10^{20}$ cm$^{-2}$, $\Gamma = 2.41\pm 0.06$
at a $\chi^2_{red} = 1.54$ for 1246 d.o.f).

\begin{figure}
\psfig{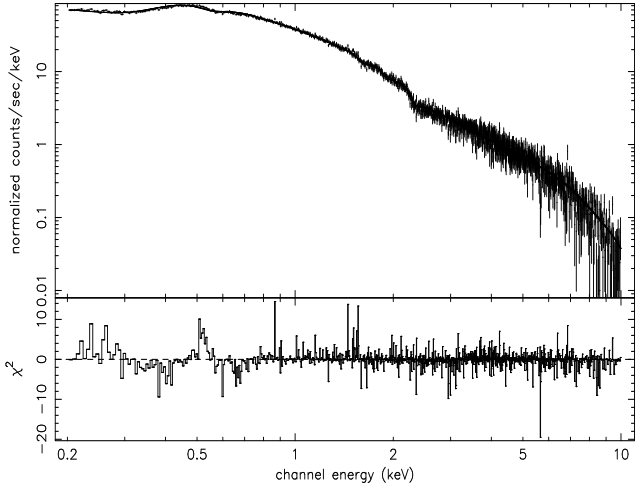}
\caption[]{\small Broken power law fit to the total PN Small Window
mode data of Mrk~421. Only single events were used.}
\label{figure:tofit}
\end{figure}

A broken power law with free absorption gives an acceptable
fit to the data (see Fig. ~\ref{figure:tofit}).
The best fit parameters are $N_H = (1.66\pm1.37)\times10^{19}$ cm$^{-2}$,
 $\Gamma_1 = 1.448\pm 0.054$, E$_{break} = (0.584\pm0.007)$ keV,
 $\Gamma_2 = 2.329\pm 0.007$
 at a $\chi^2_{red} = 1.16$ for 1244 d.o.f.
There are still systematic (instrumental) 
deviations in the residuals at low energies around 0.3~keV and 0.5~keV,
as clearly visible in Fig. ~\ref{figure:tofit}, 
leading to an inexact determination of the absorption column density.
Note the extremely high statistical significance in the data: at lower
energies every spectral bin contains more than 2000 counts!
  
With these spectral parameters the source average flux in the 2$-$6~keV
bands amounts to $(1.3 - 1.9)\times10^{-10}$ erg\,cm$^{-2}\,s^{-1}$,
where the quoted uncertainty originates mainly from the
currently uncertain contributions of the double, triple, and unrecognized
events to the 'real' source flux.

In the RGS spatial and spectral domains (i.e.  
in the dispersion vs cross-dispersion, and dispersion vs CCD pulse height,
event distributions) data cuts were made in order to extract the 
source spectra in both first
and second order. Spectra of the background (which is $\sim$ 1\% of the source
flux) were extracted from a spatial region offset from the source in the cross
dispersion direction, and as wide as that used for the source. Instrumental
responses were built for first and second order spectra for RGS1 and RGS2
(in 3000 spectral bins).

The model that best fits the \mkn RGS data is a broken power law; 
other models were tried (but no 'curved' models), but none matched the
data better.
The RGS1 first order spectra for the two exposures
were fitted simultaneously with the same absorbed broken power law model,
keeping all parameters tied, except for the normalizations which were allowed
to vary independently; the low-energy absorption column density was
fixed at the Galactic value.
The best fit parameters are 
$\Gamma_1 = 1.76 \pm 0.01$, $\Gamma_2 = 1.94 \pm 0.01$, E$_{break} = 
0.89_{-0.03}^{+0.04}$ and a reduced $\chi^2_{red}$ = 1.68.

Data from the second RGS1 exposure (strictly simultaneous with the
spectra obtained from the PN camera) were analyzed on their own, with
similar results to the total dataset.
Further, we find no evidence for any discrete spectral feature (neither in
absorption, nor in emission) in the \mkn data.
The RGS spectra
allow us to set a 90\% upper limit of 0.3 eV (0.012 \AA)
to the EW of a narrow, saturated OVII resonance absorption line at 0.574 keV
(21.6 \AA; these values correspond to 0.557 keV, and  22.3 \AA,
respectively, at the redshift of the galaxy). The presence of such an
absorption feature was  was suggested by
Guainazzi et al. (1999) from BeppoSAX observations. 
 A similar RGS upper limit applies to the EW of an OVIII
L$\alpha$ resonant absorption line at 0.654 keV.

\subsection{Spectral comparison}
The flare at the beginning of the SW mode observation was observed by
the MOS1 camera as well and could thus be fit with both, the PN
and the MOS instruments.
 
A broken power law model with the absorption fixed at the Galactic value
resulted  for the MOS1 in an acceptable fit ($\chi^2_{red}$ = 
1.04 for 213 d.o.f.) with:  $\Gamma_1 = 2.004 \pm 0.004$, 
$\Gamma_2 = 2.295 \pm 0.078$, E$_{break} = 2.33\pm0.08$ keV.
The same model (fixed Galactic absorption and broken power law) did not
yield an acceptable fit in the PN; only when the $N_H$ - value was left
as a free parameter we obtained acceptable fits with 
$\chi^2_{red} \sim 1.0$.
While the fitted  slope of the high energy power law remained stable
 ($\Gamma_2 = 2.33 \pm 0.03$) and nearly identical to the MOS1 value
in all fits, the other parameters depended strongly on the actual lower 
boundary of the energy range for the fit.
The fitted $N_H$ always remains lower than the Galactic value.
 
Similarly, broken power law fits at different intensity levels of
\mkn tend to yield nearly identical slopes at higher energies, only
the low energy slopes and the break energies seem to vary.
In conclusion, the
above fit comparisons indicate that the cross calibration of the
spectral responses still needs improving.

\section{Summary}
X-ray observations have been used frequently to 
determine the physical conditions of the central engines of BL Lac  
objects.
In most cases integration times over typically one satellite 
orbit inhibited the study of irregularities in the 
variability patterns and the puzzling spectral behavior of the sources 
on shorter time scales.
XMM$-$Newton with its  high sensitivity and broad 
energy bandwidth allows spectroscopy with unprecedented time resolution,
uninterrupted by gaps because of the long period of the satellite orbit.
     
In the extended \xmm  Cal/PV observations of \mkn in May 25, 2000 
the source was found in a relatively high state with intensity variations
by a factor of more than 3. 
In the hard energy band the source is considerably more variable 
than in the soft band,  but in the time resolved flares we 
find only a small  time lag between the hard and the soft photons.
This result and the fact that  the shape of individual pulses and their 
spectral evolution are resolvable on time scales of $\sim$ 100 secs,
puts strong constraints on time dependent radiation models for 
BL Lac jets (\cite{GEM}, \cite{KRM},\cite{CG}, Kataoka \etal 1999). 

The spectra can be well fitted by broken power laws with break energies 
around 1$-$2 keV. Flux variations seem to affect the low energy power law
slope and the value of the break energy; the higher energy slope 
appears to be quite stable.
A quantitative analysis, however, will have to await a further
reduction of the remaining calibration uncertainties. 

\vskip 0.4cm
\begin{acknowledgements}
We thank I. Papadakis for his help  with the cross-correlation analysis
 and K. Dennerl, S. Zavlin, and F. Haberl for producing off-line 
event files with currently unpublished CTE corrections.
The XMM$-$Newton project is supported by the Bundesministerium fuer
Bildung und Forschung / Deutsches Zentrum fuer Luft- und Raumfahrt 
(BMBF/DLR), the Max-Planck Society and the Heidenhain-Stiftung.
The Mullard Space Science Laboratory and Leicester University 
 acknowledge financial support
from the UK Particle Physics and Astronomy Research Council.
Part of this work was done in the TMR research network 'Accretion
onto black holes, compact stars and protostars' funded by
the European Commission under contract number ERBFMRX-CT98-0195. 
\end{acknowledgements}

\end{document}